\documentstyle[epsf]{aipproc}
\begin{document}

\title{Dilepton Production in Relativistic Heavy Ion Collisions}
\ \\
\author{Volker Koch}
\ \\
\address{Lawrence Berkeley National Laboratory\\
Berkeley, CA 94720}

\maketitle

\begin{abstract}
The present status of our understanding of low mass 
dilepton production in relativistic heavy ion 
collisions is discussed. The focus of the discussion will be the sensitivity 
of dilepton measurements to in medium changes of hadrons and the restoration
of chiral symmetry. We will finally discuss how the presence of strong
long wavelength pion modes, i.e. disoriented chiral condensates can be
seen in the dilepton spectrum.
\end{abstract}

\section*{Introduction}
Electromagnetic probes, such as photons and dileptons are especially
useful to investigate the early stage of an ultrarelativistic heavy ion
collision, since they leave the system without any final state interaction.
However, as  measurements by the DLS collaboration at the BEVALAC,
and more recently by the CERES collaboration at the CERN-SPS have
shown, a large fraction of the dilepton yield arises from the decay of
long lived states, such as the $\pi_0$, eta, or the omega. These resonances 
decay well outside the hot  and compressed region and thus  careful analysis
of the dilepton spectra is needed in order to extract the 
information about the properties of the hot and dense matter, such as
e.g. possible in medium changes of hadrons. In the low mass
region, below the phi-meson, the most important production channels are:
(i) Dalitz decays of $\eta$, $\Delta$, $\omega$, $a_1$. 
(ii) Direct decays of the vector mesons, such as $\rho$, $\omega$ and
$\Phi$. Unique to heavy-ion collisions are  rescattering
channels such as pion annihilation and bremsstrahlung due to secondary
collisions. These latter channels certainly carry  information about
the hot and dense region and due to the vector dominance formfactor  pion
annihilation  may reveal possible in medium changes of the $\rho$-meson.
(For an review of in medium changes of hadronic properties see e.g. 
\cite{KKL97}).

\section*{Dilepton production at SPS-energies}
As discussed in detail in the contribution by A. Dress \cite{Drees} dilepton
measurements for nucleus-nucleus collisions at 200 GeV per nucleon 
by the CERES collaboration 
show a considerable enhancement over the expected yield from
hadronic decays. For p+Be as well as p+Au collisions, on the other hand, the
data are consistent with the hadronic decays only. 
Certainly a large fraction of this enhancement is due pion annihilation, which
is unique to  heavy-ion experiments since they create a dense system of pions
which then can annihilate. Thus the CERES data are proof that heavy ion
collisions are more then the simple superposition of individual nucleon-nucleon
collisions and that indeed an interacting hadronic system is formed (similar
evidence is also derived from the measurement strange particle production).

\begin{figure}[htb]
\setlength{\epsfxsize=0.8\textwidth}
\centerline{\hspace{0.15\textwidth}\epsffile{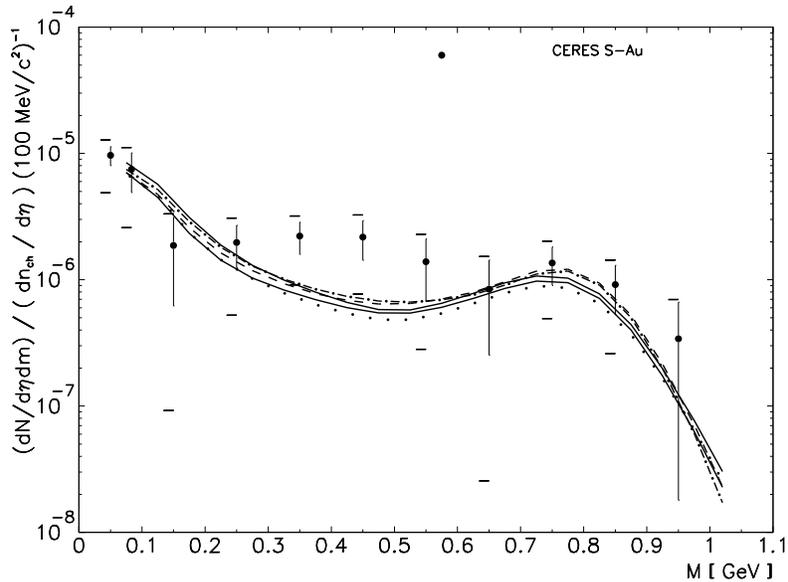}}
\caption{CERES data of S+Au in comparison with calculation based on different
initial hadronic configurations.}
\label{fig:ceres_comp}
\end{figure}

Aside from measuring in medium modifications of hadrons, dilepton measurements
may provide complementary information about the reaction dynamics and may thus
help to further specify the properties of the hadronic system generated in
these collisions. This question has been addressed in \cite{KS96}, where the
dilepton spectrum has been calculated for a large variety of initial conditions
under the constraint the the final hadronic spectra are in agreement with
experiment. Within the CERES acceptance, the variation of the resulting
dilepton mass spectra is rather small (see fig. \ref{fig:ceres_comp}). 
Thus  the measurement of an dilepton invariant mass spectrum is unlikely 
to further specify the configuration of the hadronic phase. On the positive 
side this result shows that large deviations of the data
from the hadronic calculation cannot simply be attributed to the lack of
knowledge of the specific configuration of the hadronic phase. 
And indeed, as compared with the central points of the CERES measurement for
$S+Au$, there is a considerable deviation at invariant masses of about $400 \,
\rm MeV$. However, one should also point out, that within the
systematic and statistical error quoted by the CERES collaboration, the data
can be understood by simply including pion annihilation without any further in
medium modifications. 
However, if one seeks to reproduce the central values of the CERES data,
certain in medium modifications have to be included. 
One is  follow the conjecture of Brown and Rho \cite{BR91} 
that the mass of the  $\rho$ 
meson scales with the quark condensate, which is expected to be reduced at
the densities and temperatures reached in these collisions. As a consequence
the mass of the $\rho$ drops, providing more strength in the low mass region
of the dilepton spectrum. Following this prescription, Li et al. can reproduce
the central points of the CERES measurements \cite{LKB96}. 
However, at least at low temperatures, the Brown-Rho
scaling hypothesis can be ruled out by simple current algebra arguments 
\cite{DEI90}, which shows that to order $T^2$ the mass of the $\rho$ does not
change while to this order  the quark condensate drops. 

More conventional calculations of in medium effects on the dilepton production
determine the properties of the $\rho$-meson or more precisely 
the current-current correlator in a system of pions and nucleons/deltas. 
So far essentially three
different in medium correction have been considered, which are schematically
depicted in fig. 2.
\begin{figure}[htb]
\setlength{\epsfxsize=0.5\textwidth}
\centerline{\hspace{0.15\textwidth}\epsffile{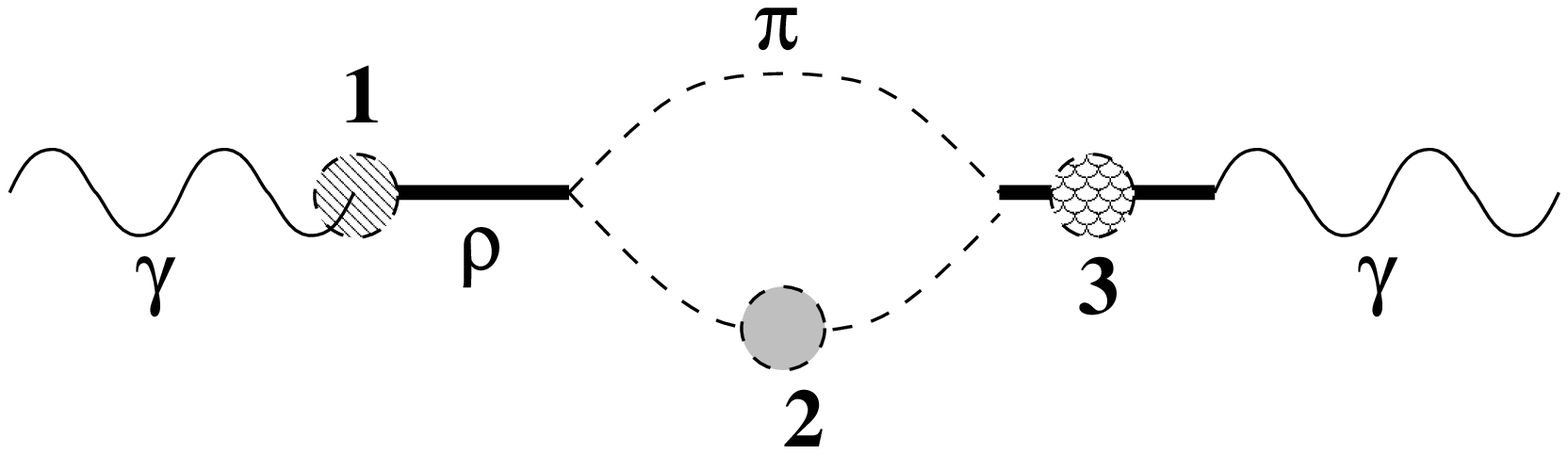}}
\end{figure}

(1) The $\rho \gamma$ coupling is screened due to pion loops \cite{SLK95,SK96}
This effect is a direct consequence of the partial restoration of chiral
symmetry and it reduces the strength below the $\rho$ peak. \\
(2) If one understands the $\rho$ as a $\pi-\pi$ resonance, its properties are
    changed due to in medium modifications of the pions. These include
a change of the pion dispersion relation due to thermal pions \cite{KS96,SKL96}
and due to the coupling to delta-hole and nucleon-hole states 
\cite{Rapp,Weise,Steele} once  baryons are taken into account. 
These states also give rise to additional inelasticities at low invariant
resulting in an increased strength around 400 MeV in the 
imaginary part of the current-current correlator. One should note, however, 
that to leading order in the density, these contributions are nothing else but
typical bremsstrahlung diagrams. 
Furthermore, these  additional inelasticities at low
masses seem to be sufficient to saturate the QCD-sum rules \cite{Weise},
without an explicit change in the mass of the $\rho$ meson. \\
(3) The $\rho$ can also couple to $N^*(1720) $-hole states. This effect,
    originally proposed by Friman and Pirner \cite{FP97} leads to a softening
    of the dispersion relation of the $\rho$ meson and provides additional
    strength in the dilepton spectrum at low invariant masses. Since the
    coupling to the $N^*(1720) $-hole state is p-wave, only dileptons
    with finite momentum with respect to the matter restframe are enhanced.

A combination of effects (2) and (3) seems leads to an improved description of
the CERES data \cite{Rapp}.

\section*{Dileptons from DCC-states}
The restoration of chiral symmetry in relativistic heavy ion collisions can,
under certain circumstances, lead to a strong enhancement of low momentum pion
modes which form a so called disoriented chiral condensate (DCC) 
\cite{Bjorken,Wilckzek}. So far, proposed observables which are sensitive to 
these DCC states have been in the pion sector only, where strong final state
interactions may destroy them. However, the presence of a DCC state also leads
to a strong and unique signal in the dilepton channel. Assume a thermal
pion annihilates with a pion from a DCC. Since the DCC represents a large
phase space density localized at small momenta, one would expect that this
phase space distribution is reflected in the dilepton invariant mass as well as
momentum spectrum. This is indeed the case as one can see in figures 3 and 4. 
(for details see \cite{KKRW97}). In fig. 3 we show the resulting invariant 
mass distribution for thermal initial conditions and for so called quench 
initial conditions; the latter lead to the formation of DCC-states. The
resulting momentum spectrum for an invariant mass of $M = 300 \, \rm MeV$ is 
shown in fig. 4. Clearly a strong enhancement (about a factor of 100) can be
seen close to twice the pion mass. The enhancement is localized in invariant
mass as well as in momentum, reflecting the localized phase space distribution
of the DCC-state.
\begin{figure}[htb]
\setlength{\epsfxsize=\textwidth}
\centerline{\epsffile{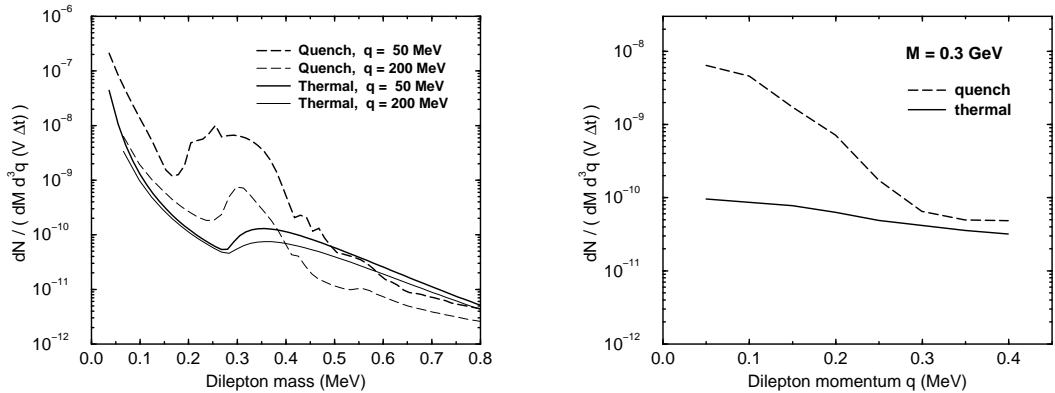}}
\caption{Dilepton invariant mass (left) and momentum (right) 
spectra for thermal (full lines) and quench
(dashed lines) initial conditions.  The mometum spectrum is for an invariant
mass of $M= 300 \, \rm MeV$.}
\end{figure}
Since this enhancement is confined to momenta below 300 MeV,
it does not affect the CERES measurement, where an acceptance cut
of $p_t \geq 200 \, \rm MeV$ for each individual dilepton is imposed. However,
if this cut could be relaxed to $p_t \geq 100 \, \rm MeV$ a factor of 10
enhancement in the invariant mass spectrum should be visible. \\

Acknowledgments: I would like to thank Y. Kluger, J. Randrup, C. Song, and
X.N. Wang, whom I have collaborated with on the topics discussed here. 
This work was supported by the Director, 
Office of Energy Research, Office of High Energy and Nuclear Physics, 
Division of Nuclear Physics, and by the Office of Basic Energy
Sciences, Division of Nuclear Sciences, of the U.S. Department of Energy 
under Contract No. DE-AC03-76SF00098.

\end{document}